\documentclass[prl,showpacs,nofootinbib,preprintnumbers,twocolumn,floatfix]{revtex4}

\usepackage{graphicx}
\usepackage{amssymb}
\usepackage{pifont}

\def\lQ{\Lambda_{\rm QCD}}
\newcommand{\be}{\begin{equation}}{\bf }
\newcommand{\ee}{\end{equation}}
\newcommand{\bea}{\begin{eqnarray}}
\newcommand{\eea}{\end{eqnarray}}
\newcommand{\nn}{\nonumber}

\def\siml{{\ \lower-1.2pt\vbox{\hbox{\rlap{$\langle$}\lower6pt\vbox{\hbox{$\sim$}}}}\ }}

\begin{document}

\title{Effective string theory constraints on the long distance behavior of the 
subleading
potentials}
\author{Guillem P\'erez-Nadal$^1$}
\author{Joan Soto$^2$}
\affiliation{$^1$ Departament de F\'isica Fonamental\\
$^2$ Departament d'Estructura i Constituents de la Mat\`eria \\and \\Institut de Ci\`encies del Cosmos\\
 Universitat de Barcelona\\
Diagonal 647, 08028 Barcelona, Catalonia, Spain}
\preprint{UB-ECM-PF 08/20}

\begin{abstract}

The dynamics of heavy quarkonium systems in the strong coupling regime reduces to a quantum mechanical problem with a number of potentials which may be organized in powers of $1/m$, $m$ being the heavy quark mass. The potentials must be calculated non-perturbatively, for instance in lattice QCD. 
It is well known that the long distance behavior of the static ($1/m^0$) potential is well reproduced by an effective string theory. We show that this effective string theory, if correct, should also reproduce the long distance behavior of all $1/m$ suppressed potentials. We demonstrate the practical usefulness of this result by finding a suitable parameterization of the recently calculated $1/m$ potential. We also 
calculate the $1/m^2$ velocity dependent and spin dependent potentials. Once Poincar\'e invariance is implemented, the shapes of most of the spin independent potentials are fully predicted in terms of the string tension, and the shapes of the spin dependent ones in terms of a single parameter.
\end{abstract}

\maketitle

Heavy quarkonium systems have played a major role in our understanding of QCD
(see \cite{Brambilla:2004wf} for a review). 
The early successes of non-relativistic potential models in describing the gross features of the spectrum, can nowadays be understood as emanating from QCD in a particular kinematical regime.
The heavy quarks in the heavy quarkonium rest frame move slowly, with a velocity $v\ll 1$, which generates a hierarchy of physical scales  $m \gg mv \gg mv^2$ ($1/mv$ is the typical size of the system and $mv^2$ the typical binding energy) in addition to $\lQ$, the typical hadronic scale. This hierarchy is most conveniently exploited using the effective field theories (EFT) of NRQCD \cite{Caswell:1985ui,Bodwin:1994jh} and pNRQCD \cite{Pineda:1997bj,Brambilla:1999xf}, which are built in such a way that they are equivalent to QCD in the kinematical regime they hold (see \cite{Brambilla:2004jw} for a review). It was shown in \cite{Brambilla:1999xf} that in the case $mv \sim \lQ$ the relevant degrees of freedom of pNRQCD (and hence of QCD) reduce to those of non-relativistic potential models. The potentials to be input in pNRQCD, however, have precise formulas in terms of objects computable from QCD. Some of these formulas were known since long \cite{Brown:1979ya,Eichten:1980mw,Barchielli:1986zs}, but others were uncovered when formulating this problem in the EFT framework, like the $1/m$ potential \cite{Brambilla:2000gk}. 

The potentials have been computed on the lattice with increasing precision \cite{Bali:1992ab,Bali:1997am,Necco:2001xg,Koma:2006si,Koma:2006fw} . Convenient and economical parameterizations of lattice data are necessary in order to include the potentials as simple functions in the Schr\"odinger equation. For the static potential the na\"\i ve addition of the short distance one-gluon exchange potential and the long distance linear potential, as predicted by the effective string theory (EST) \cite{Nambu:1978bd}, which is known as the Cornell potential \cite{Eichten:1978tg}, provides a good description of lattice data and has been very successful in phenomenological applications. Corrections to the long distance linear behavior can be calculated in a systematic manner in the EST \cite{Luscher:1980fr,Luscher:2002qv} (see also \cite{Polchinski:1991ax}). For the subleading potentials, so far the only constraint which has been used for such parameterizations, is that at short distances, the potentials must approach their perturbative expressions. The long distance behavior has traditionally been a matter of guess work, being quite common the use of polynomials in $1/r$ (lately powers of $r$ have also been used). The aim of this letter is to show that the EST also predicts the long distance behavior of the $1/m$ suppressed potentials, and hence may become an extremely useful tool in order to find suitable parameterizations of lattice data.

The static potential can be obtained from the vacuum expectation value of the rectangular Wilson loop $W(T,r)$ \cite{Wilson:1974sk}. The EST hypothesis maintains that at long distances ($r\Lambda_{QCD} \gg 1$) this expectation value can be obtained from a string action,
\be 
\lim_{T\rightarrow \infty} \langle 0| W(T,r)|0\rangle=Z\int {\cal D} \xi^l e^{iS_{\rm string}(\xi^l)}
\label{string}
\ee
where $Z$ is an unknown constant, and $\xi^l=\xi^l (t,z)$, $l=1,2$, are the transverse components of the string, which fulfill the boundary conditions $\xi^l (t,r/2)=\xi^l (t,-r/2)=0$. The string action may be written as \cite{Luscher:2002qv}
\be
S_{\rm string}=-{\kappa}\int dt\,dz\,\left( 1-{1\over 2}\partial_\mu\xi^l\partial^\mu\xi^l\right)
\label{action}
\ee
where $\kappa$ is the string tension.
This action is corrected by higher order terms in the EST counting, and can be obtained as a long wavelength limit of the Nambu-Goto action. 
Equations (\ref{string}) and (\ref{action}) give rise to the following prediction for the long distance behavior of the static potential \cite{Luscher:1980fr},
\be
V^{(0)}(r)=\kappa r+\mu-\frac{\pi}{12 r}
\label{v0}
\ee
where $\mu$ is an unknown constant. This result agrees with lattice data for $\kappa\simeq 0.21$ GeV$^2$ \cite{Luscher:2002qv}.

The $1/m$ suppressed potentials are given by expectation values of suitable operator insertions in the rectangular Wilson loop (see \cite{Brambilla:2000gk,Pineda:2000sz} for concrete formulas). 
Since the large distance behavior of the expectation value of the Wilson loop is given by an EST, it is natural to expect that the suitable operator insertions that the $1/m$ potentials need also have a representation in the EST. 
In order to pin down the mapping it is convenient to express the operator insertions in a gauge invariant fashion. This is achieved by introducing two spinless (Grassmann) fields $\psi$ and $\chi$. $\psi$ annihilates a static source in the fundamental representation at the point ${\bf r}/2=(0,0,r/2)$ and $\chi$ creates a static source in the anti-fundamental representation at the point $-{\bf r}/2$, $\{\psi^\dagger , \psi \}=\{\chi^\dagger , \chi \}=1$, the remaining fixed-time (anti-)commutators being zero. The QCD Lagrangian is then augmented by
\bea
\delta L_{QCD}&=&\psi^\dagger (t)\left( i\partial_0-gA_0(t,{\bf r}/2)\right) \psi (t) +\nn\\
&& \chi^\dagger (t)\left( i\partial_0-gA_0(t,-{\bf r}/2)\right) \chi (t)
\eea
The expectation value of the rectangular Wilson loop $W(T,r)$ can be rewritten as 
\be
\langle 0| W(T,r)|0\rangle=
\langle 0|
O({T\over 2},{\bf r})O^\dagger(-{T\over 2},{\bf r})
|0\rangle 
\label{wloop}
\ee
\be
 O(t,{\bf r})= \chi^\dagger (t)\phi (t,-{{\bf r}\over 2};t,{{\bf r}\over 2})\psi (t)
\ee
$\phi (t,{\bf r};t,{\bf r}')$ is the straight Wilson line joining the points ${\bf r}$ and ${\bf r}'$ at the time $t$. In this formalism the 
insertions of chromoelectric and chromomagnetic operators (see (\ref{1m}) below and ref. \cite{Pineda:2000sz}) correspond to insertions in (\ref{wloop}) of the following gauge invariant operators,
\bea
\psi^\dagger (t) {\bf E}^i(t,{{\bf r}\over 2})\psi (t) &,& \psi^\dagger (t) {\bf B}^i(t,{{\bf r}\over 2})\psi (t)\nn\\
-\chi^\dagger (t) {\bf E}^i(t,-{{\bf r}\over 2})\chi (t) &,& -\chi^\dagger (t) {\bf B}^i(t,-{{\bf r}\over 2})\chi (t) 
\label{NRQCDoperators}
\eea
For instance, let us denote as $\langle{\bf E}^i(t,{\bf r}/2)$ $ {\bf E}^i(t',{\bf r}/2)\rangle$ the expectation value of the insertions of two chromoelectric fields at the points $(t,{\bf r}/2)$ and $(t',{\bf r}/2)$ of the Wilson loop ($T/2 > t > t' > -T/2$). We have,
\bea
&&\langle {\bf E}^i(t,{\bf r}/2) {\bf E}^i(t',{\bf r}/2) \rangle =\nn\\
&& \langle 0\vert 
O({T\over 2},{\bf r})\psi^\dagger (t) {\bf E}^i(t,{{\bf r}\over 2})\psi (t) \\
&& \times \psi^\dagger (t') {\bf E}^i(t',{{\bf r}\over 2})\psi (t')O^\dagger(-{T\over 2},{\bf r}) \vert 0 \rangle\nn
\label{insert}
\eea
This way of rewriting the operator insertions in the Wilson loop is especially convenient for the mapping into the EST. In the limit $T\to\infty$, which is taken in the computation of the $1/m$ suppressed potentials, the chromoelectric and chromomagnetic insertions reduce to correlation functions of the gauge invariant operators (\ref{NRQCDoperators}). These correlation functions can now be mapped into the EST as correlation functions of some suitable EST operators. 

Therefore, what we have to do is to find a representation of operators like (\ref{NRQCDoperators}) in terms of string variables, under the guidance of the global symmetries of the system. The latter correspond to the $D_{h\infty}$ group, the symmetries of a diatomic molecule (changing P by CP), and time reversal. In order to identify the implementation of the symmetry in the EST, it is convenient to choose a worldsheet parameterization in which evolution is described by time, the zeroth coordinate of the string, and the labeling by the z coordinate, the last coordinate of the string, as it has already been implemented in (\ref{action}). 
For the building blocks of (\ref{NRQCDoperators}), we have the following transformation properties with respect to the generators of $D_{h\infty}$ (${\bf z}=(0,0,z)$):
\begin{itemize}
\item Rotations with respect to the z-axis
\bea
&& {\bf E}^i(t,{\bf z})\to R^{ij} {\bf E}^j(t,{\bf z})\nonumber\\
&&{\bf B}^i(t,{\bf z})\to R^{ij} {\bf B}^j(t,{\bf z})\nn\\
&& \psi (t) \to \psi (t) 
\; ,\; \chi (t) \to \chi (t)
\eea
\item Reflection with respect to the zx-plane
\bea
&&{\bf E}^i(t,{\bf z})\to \rho^{ij}{\bf E}^j(t,{\bf z})\nonumber\\
&&{\bf B}^i(t,{\bf z})\to -\rho^{ij}{\bf B}^j(t,{\bf z})\nn\\
&& \psi (t) \to \psi (t) 
\; ,\; \chi (t) \to \chi (t)
\eea
\item CP
\bea
&&{\bf E}^i(t,{\bf z})\to \left({\bf E}^i\right)^{T}(t,-{\bf z})\nonumber\\
&&{\bf B}^i(t,{\bf z})\to -\left({\bf B}^i\right)^{T}(t,-{\bf z})\nn\\
&& \psi (t) \to \chi^\ast (t) 
\; ,\; \chi (t) \to \psi^\ast (t)
\eea
\end{itemize}
Under time reversal they transform as follows:
\begin{itemize}
\item T
\bea
&&{\bf E}^i(t,{\bf z})\to {\bf E}^i(-t,{\bf z})\nonumber\\
&&{\bf B}^i(t,{\bf z})\to -{\bf B}^i(-t,{\bf z})\nn\\
&& \psi (t) \to \psi (-t) 
\; ,\; \chi (t) \to \chi (-t)
\eea
\end{itemize}
In these equations, $R^{ij}$ is the rotation matrix, $\rho^{ij}={\rm diag}(1,-1,1)$, and $T$ stands for transpose (with respect to color indices). On the string theory side, the building blocks, namely the string coordinates $\xi^i(t,z)$ (with $\xi^3=z$), 
transform as follows:
\begin{itemize}
\item Rotations with respect to the z-axis
\bea
&& \xi^i(t,z) \to R^{ij}\xi^j(t,z) 
\eea
\item Reflection with respect to the zx-plane
\bea
&&\xi^i(t,z)\to \rho^{ij}\xi^j(t,z) 
\eea
\item CP
\bea
&&\xi^i(t,z)\to -\xi^i(t,-z)
\eea
\item T
\bea
&&\xi^i(t,z)\to \xi^i(-t,z)
\eea
\end{itemize}
We find that the following mapping satisfies the symmetry requirements,
\bea
\psi^\dagger (t) {\bf E}^l(t,{{\bf r}\over 2})\psi (t) &\mapsto &\Lambda^2\partial_z\xi^l (t,{r\over 2})\label{mapping}\nn\\
\chi^\dagger (t) {\bf E}^l(t,-{{\bf r}\over 2})\chi (t) &\mapsto &-\Lambda^2\partial_z\xi^l (t,-{r\over 2})\nonumber\\
\psi^\dagger (t) {\bf B}^l(t,{{\bf r}\over 2})\psi (t)&\mapsto &\Lambda'\epsilon^{lm}\partial_t\partial_z\xi^m(t,{r\over 2})\nn\\
\chi^\dagger (t) {\bf B}^l(t,-{{\bf r}\over 2})\chi (t)&\mapsto & \Lambda'\epsilon^{lm}\partial_t\partial_z\xi^m(t,-{r\over 2})\nn\\
\psi^\dagger (t) {\bf E}^3(t,{{\bf r}\over 2})\psi (t) &\mapsto &{\Lambda''}^2\\
\chi^\dagger (t) {\bf E}^3(t,-{{\bf r}\over 2})\chi (t) &\mapsto &-{\Lambda''}^2\nonumber\\
\psi^\dagger (t) {\bf B}^3(t,{{\bf r}\over 2})\psi (t)&\mapsto &{\Lambda'''}\epsilon^{lm}\partial_t\partial_z\xi^l(t,{r\over 2})\partial_z\xi^m(t,{r\over 2})\nn\\
\chi^\dagger (t) {\bf B}^3(t,-{{\bf r}\over 2})\chi (t)&\mapsto & {\Lambda'''}\epsilon^{lm}\partial_t\partial_z\xi^l(t,-{r\over 2})\partial_z\xi^m(t,-{r\over 2})\nn
\eea
where $l,m = 1,2$ and $\Lambda$, $\Lambda'$, $\Lambda''$, $\Lambda''' \sim \lQ$ are unknown constants with dimension of mass. The assignment above agrees with the early assignment in ref. \cite{Kogut:1981gm}. The EST provides an expansion of the physical observables in terms of $1/r\lQ$, transverse string coordinates must be counted as $1/\lQ$, whereas $\partial_z$ and $\partial_0$ like $1/r$. Hence the expressions in (\ref{mapping}) will be corrected by higher order operators in the EST counting. The expression of the $1/m$ potentials in the EST will be obtained by substituting the operators on the lhs of (\ref{mapping}) by the operators on the rhs of (\ref{mapping}) and calculating the expectation values with the EST action (\ref{action}).

Let us illustrate it by calculating the EST expression of the $1/m$ potential. For this potential we have \cite{Pineda:2000sz} 
\be
V^{(1,0)}(r)=-{g^2\over 2}\int_{0}^\infty dt\, t\,\langle\!\langle {\bf E}^i(t,{{\bf r}\over 2}){\bf E}^{i}(0,{{\bf r}\over 2})\rangle\!\rangle_c
\label{1m}
\ee
where $\langle\!\langle \cdots \rangle\!\rangle$ means that the expectation value of the operator insertions in the Wilson loop (e.g. (8))
is normalized to the expectation value of the Wilson loop (\ref{wloop}), and the subscript $c$ stands for connected. Hence the EST representation is 
\be
V^{(1,0)}(r)=-{g^2\Lambda ^4\over 2}\int_{0}^\infty dt\, t\,\partial_z\partial_{z'}G_{F}^{ll}(t,{r\over 2};0,{r\over 2})
\label{1ms}
\ee
where $G_{F}^{lm}(t,z;t',z')=\langle \xi^l(t,z)\xi^m(t',z')\rangle$. This integral is most easily computed by performing a Wick rotation to imaginary time. For the calculation of the correlator we obtain
\bea
&&G_{F}^{lm}(it ,z;it',z')=\delta^{lm}\frac{1}{4\pi\kappa}\times\nonumber\\
&&\ln\left\{\frac{\cosh\left[\frac{\pi}{r}(t-t')\right]+\cos\left[\frac{\pi}{r}(z+z')\right]}{\cosh\left[\frac{\pi}{r}(t-t')\right]-\cos\left[\frac{\pi}{r}(z-z')\right]}\right\}
\label{correlator}
\eea 
The time integration in (\ref{1ms}) suffers from an UV divergence, which may be regulated by introducing a cut-off for small times. The contribution from this cut-off is just an additive constant to the potential, which may be absorbed into the additive constant that appears in the EST result for the static potential (\ref{v0}). Up to a constant term, we then obtain
\be
V^{(1,0)}(r)=\frac{g^2\Lambda^4}{\pi\kappa}\,\ln \left(\sqrt\kappa\, r\right)
\label{log}
\ee
Hence we obtain the non-trivial result that the $1/m$ potential must grow logarithmically at large $r$. Let us compare this result with available lattice data. We fitted a curve of the form $V^{(1,0)}(r)=a\log r+b$ to the data in \cite{Koma:2007jq} at $\beta=6/g^2=6.00$ and $r>0.2$ fm. Note that this range already corresponds to the intermediate and long distance regimes $r\gtrsim\lQ^{-1}$. The result is plotted in Fig. 1.
\begin{figure}
\centering
\includegraphics[angle=-90,scale=0.39]{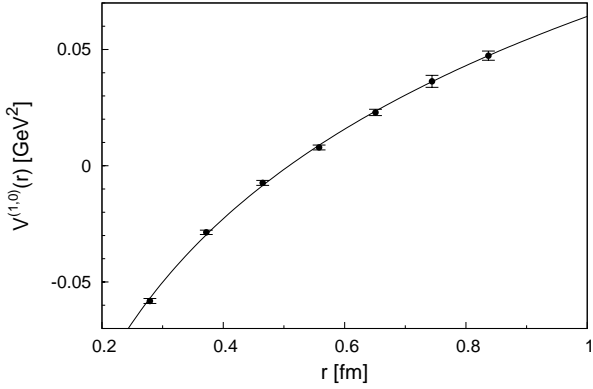}
\caption{The lattice data for $V^{(1,0)}(r)$, fitted to the EST prediction $V^{(1,0)}(r)=a\log r+b$.}
\end{figure}
As we can see, the fit is very good, with a reduced chi-square $\chi^2/N_{df}=0.93$
\footnote{We have considered the errors of the different lattice points uncorrelated. When the correlations are taken into account $\chi^2/N_{df}$ becomes larger but still of order one \cite{miho}.}. Of course, for phenomenological applications (see for instance \cite{DomenechGarret:2008vk}) a short distance piece compatible with perturbation theory ($\sim 1/r^2$) should be ``added'' to the long distance behavior above.

Some of the $1/m^2$ potentials are related to the correlator (\ref{correlator}), and hence can be easily obtained from it,
\bea
&&V^{(2,0)}_{p^2}(r)=V^{(1,1)}_{p^2}(r)=0\nn\\
&&V^{(2,0)}_{L^2}(r)=-V^{(1,1)}_{L^2}(r)=-\frac{g^2\Lambda^4}{6\kappa}r
\label{1m2}
\eea
The velocity dependent potentials $V^{(2,0)}_{p^2}(r)$ and  $V^{(1,1)}_{p^2}(r)$ may receive non-vanishing contributions at NNLO, and, hence, up to logarithmic corrections, they are expected to scale as $V^{(2,0)}_{p^2}(r)\sim V^{(1,1)}_{p^2}(r)\sim C/r$ ($V^{(2,0)}_{p^2}(r)$ may develop a constant piece due to a contact term, similar to the ones appearing in (\ref{contact}) below). We obtain from (\ref{1m2}) the following model-independent predictions for the long range behavior of these potentials,
\bea
 {V^{(2,0)}_{L^2}(r)\over  V^{(1,1)}_{L^2}(r)}  =  -1 & , &
 {r^2 {d \over d r} V^{(1,0)}(r)\over  V^{(2,0)}_{L^2}(r)}  = -{6\over\pi}
\eea

Let us next turn to the potentials involving chromomagnetic fields (spin dependent potentials). We obtain for the spin-orbit potentials,
\bea
V_{LS}^{(2,0)}(r)&=&-{\mu_{c}^2\over r}-{g^2c_{F}^{(1)}\Lambda'\Lambda^2\over\kappa r^2} \nn\\ 
V_{L_2S_1}^{(1,1)}(r) &=&-{g^2c_{F}^{(1)}\Lambda'\Lambda^2\over\kappa r^2}
\label{so}
\eea
where $c_{F}^{(1)}$ is a matching coefficient of the NRQCD Lagrangian, which is inherited by
the spin-orbit interaction (see \cite{Brambilla:2004jw}). $V_{LS}^{(2,0)}(r)$ is UV divergent and requires regularization and renormalization. This is not a problem of the EST itself but rather one inherited from the static limit of QCD. The introduction of the static fields $\psi (t)$ and $\chi (t)$ makes the solution of the problem straightforward. Indeed, whenever we have a time ordered product of local operators, contact (local) terms of dimension equal or smaller than the sum of the dimensions of the operators must generically be added in order to obtain finite results. In the case of $V_{LS}^{(2,0)}(r)$, which involves the time ordered product
\be
\epsilon^{lm}\psi^\dagger (t) {\bf B}^l (t, {{\bf r}\over 2})\psi (t)\psi^\dagger (0) {\bf E}^m (0, {{\bf r}\over 2})\psi (0)
\ee
only the following terms 
are possible \footnote{Note that $\psi^\dagger (0)\psi (0)$ is the identity operator in the subspace spanned by $\psi^\dagger (0)$, and hence operators involving higher powers of it are redundant.},
\bea
&& \epsilon^{lm}\psi^\dagger (t) {\bf B}^l (t, {{\bf r}\over 2})\psi (t)\psi^\dagger (0) {\bf E}^m (0, {{\bf r}\over 2})\psi (0)\longrightarrow \nn\\
&& \epsilon^{lm}\psi^\dagger (t) {\bf B}^l (t, {{\bf r}\over 2})\psi (t)\psi^\dagger (0) {\bf E}^m (0, {{\bf r}\over 2})\psi (0)
\label{contact}\nn\\
&& + i\left( c_1 \delta' (t) +  c_2 \delta''' (t) \right)\psi^\dagger (0)\psi (0)\\
&& + \; i c_3 \delta' (t) \psi^\dagger (0){\bf E}^3 (0, {{\bf r}\over 2})\psi (0) \nn
\eea
$c_1$, $c_2$ and $c_3$ are real constants. The term with $c_2$ is subleading in the EST counting, 
but $c_1$ and $c_3$ are  not.
We use the same regularization as for $V^{(1,0)}(r)$ and add a suitable contact term corresponding to the EST representation of the terms with $c_1$ and $c_3$ in (\ref{contact}), which turn out to be proportional to the identity operator, in order to make the final expression finite. The coefficient $\mu_{c}^2$ appearing in (\ref{so}) depends on the finite piece of this contact term and must be considered an additional free parameter. For the spin-spin potentials we get zero at LO, which is consistent with the argument put forward in \cite{Kogut:1981gm}. However, at NLO they might receive non-vanishing contributions. Up to logarithmic corrections, we expect them to scale as $V_{S^2}^{(1,1)}\sim V_{{\bf S}_{12}}^{(1,1)}\sim C/r^5$, which may explain the sharp drop observed in lattice calculations \cite{Koma:2006fw}. Note that these contributions would be $m^2/\lQ^2$ enhanced with respect to the one found in \cite{Kogut:1981gm}.    

Before closing, it is interesting to explore the constraints that Poincar\'e invariance imposes on the potentials \cite{Brambilla:2001xk} with regard to the EST results above. The Gromes relation \cite{Gromes:1984ma}  and the first BBMP relation \cite{Barchielli:1988zp} fix $\mu_{c}^2$ in (\ref{so}) and $\Lambda^2$ to
\bea
\mu_{c}^2=\kappa /2 & , &
g\Lambda^2=\kappa
\eea
The other BBMP relations are satisfied without any further constraints. This is a remarkable result. It fixes the coefficients of the $1/m$ potential and of the velocity dependent potentials in terms of the slope of the static potential (the string tension $\kappa$). For the $1/m$ potential the fit value of the coefficient $a=0.095$ GeV$^2$, whereas the previous relation gives $a=\kappa/\pi=0.067$ GeV$^2$. The difference may be due to two reasons: (i) the lattice data of \cite{Koma:2007jq} are not in the continuum and, hence, small violations of Poincar\'e invariance are expected, and (ii) higher order terms in the EST, which have not been considered, the most important of which goes like $C/r^2$,  up to logarithms.

In summary, we have shown how EST can be used to extract the long distance part of the $1/m$ suppressed potentials. As an example, we have quantitatively compared with lattice data in the case of the $1/m$ potential and have found an excellent agreement. We expect a similar agreement for the remaining potentials. When Poincar\'e invariance is used, the shapes of the spin-independent potentials are fully predicted (at LO in the EST expansion), and the shapes of the spin-dependent ones are given in terms of a single parameter.

We believe our results are important  from two different points of view. On the one hand, we have obtained for the first time a satisfactory parameterization of the $1/m$ potential at long distances, which can now be used to compute the $1/m$ correction to the heavy quarkonium spectrum. On the other hand, there is no available proof of the idea that QCD is equivalent to EST at long distances. Our results provide a number of new ways to test whether this idea is valid or not.

\begin{acknowledgments} 
We thank Gunnar Bali for bringing to our attention ref. \cite{Kogut:1981gm} and Miho Koma for making available to us the lattice data of refs. \cite{Koma:2006si,Koma:2007jq}. We acknowledge financial support from the RTN Flavianet MRTN-CT-2006-035482 (EU), the  FPA2007-60275/, FPA2007-66665-C02-01/, FPA2007-66665C02-02/ MEC grants, and  CPAN CSD2007-00042 (Spain), and the 2005SGR00564 and 2005SGR00082 CIRIT grants (Catalonia).

\end{acknowledgments}

\end{document}